\documentclass[xcolor=pdftex,usenames,dvipsnames,table]{PoS}

\usepackage{slashed}
\usepackage{graphicx}
\usepackage{amssymb,amsmath} \usepackage{pifont}
\usepackage{multirow,lscape}
\usepackage{array}
\usepackage[lofdepth,lotdepth]{subfig}
\usepackage{verbatim}
\usepackage{bm}
\usepackage{comment}

\graphicspath{{./figs/}}

\providecommand{\abs}[1]{\lvert#1\rvert}

\providecommand{\matrixe}[3]{\langle#1\lvert#2\rvert#3\rangle}

\providecommand{\mr}[1]{\mathrm{#1}}
\providecommand{\mc}[1]{\mathcal{#1}}
\providecommand{\ek}{\ensuremath{\abs{\varepsilon_K}}}
\providecommand{\vcb}{\ensuremath{\abs{V_{cb}}}}
\providecommand{\dps}{\ensuremath{\bar{B}\to D^{(*)}\ell\bar{\nu}}}
\providecommand{\ds}{\ensuremath{\bar{B}\to D^*\ell\bar{\nu}}}
\providecommand{\CL}{\nonumber\\}
\newcolumntype{C}[1]{>{\centering\arraybackslash}p{#1}}
\definecolor{dkgray}{RGB}{145,145,145}
\definecolor{violet}{RGB}{50,0,200}
\newcommand{\com}[1]{}

\title{%
Improved currents for \dps\ form factors from Oktay-Kronfeld heavy quarks
}

\ShortTitle{Improved currents for heavy quarks}

\author{%
  \speaker{Jon A.~Bailey}, Yong-Chull Jang, Weonjong Lee, Jaehoon Leem \\
  Lattice Gauge Theory Research Center, CTP, and FPRD \\
  Department of Physics and Astronomy \\ 
  Seoul National University,
  Seoul, 151-747, South Korea\\
  E-mail: \email{jonbailey@snu.ac.kr}, \email{wlee@snu.ac.kr}
}

\author{SWME Collaboration}

\abstract{%
The CKM matrix element \vcb\ can be extracted by combining experimentally
determined branching fractions for \dps\ decays with form
factors from the lattice.  While successful, the precision of this approach has
been limited by heavy-quark discretization effects.  An improved version of the
Fermilab action, the Oktay-Kronfeld action, can be used to reduce heavy-quark
discretization effects in calculations performed at the physical bottom and
charm quark masses.  Treating charm and bottom quarks as massive, we are
carrying out improvement of the flavor-changing currents
through third order in the momentum (HQET) expansion.
}

\FullConference{The 32nd International Symposium on Lattice Field Theory\\
                 23-28 June, 2014\\
                 Columbia University, New York, NY}

\begin{document}

\section{\label{sec:intro}Introduction}
The CKM matrix element \vcb\ enters searches for new physics in the quark
flavor sector of the Standard Model (SM).  Parametric uncertainties from \vcb\
dominate uncertainties in SM calculations of the branching ratios for the rare
decays $K_L\to\pi^0\nu\bar{\nu}$ and $B_{(s)}^0\to\mu^{+}\mu^{-}$, prime candidates
for new physics, as well as SM calculations of the indirect CP violation
parameter \ek, which provides an input to the global unitarity triangle
analysis.  

The exclusive semileptonic decays \dps\ proceed at rates proportional
to
$\vcb^2$~\cite{Manohar:2000dt,Sirlin:1981ie,Ginsberg:1968pz,
Ginsberg:1969jh,Atwood:1989em}.
\begin{align}
\frac{d\Gamma}{d\omega}(\bar{B}\to D\ell\bar{\nu}) &= \frac{G_F^2 \vcb^2 M_B^5}{48\pi^3}(\omega^2-1)^{3/2}r^3(1+r)^2F_D^2(\omega)\,,\label{eq:b2d} \\
\frac{d\Gamma}{d\omega}(\bar{B}\to D^*\ell\bar{\nu}) &= \frac{G_F^2 \vcb^2 M_B^5}{4\pi^3}\abs{\eta_{EW}}^2(1+\pi\alpha)(\omega^2-1)^{1/2}r^{*3}(1-r^*)^2\chi(\omega)F_{D^*}^2(\omega)\,,\label{eq:b2dstar}
\end{align}
where $\omega = v_B\cdot v_{D^{(*)}}$ is the velocity transfer
(proportional to the $D^{(*)}$ recoil energy, in the $B$ rest frame),
and $r^{(*)} = M_{D^{(*)}}/M_B$ is the ratio of the parent to the daughter
meson mass.
$\pi\alpha$ and $\eta_{EW}$ are higher order electroweak corrections.
$\pi\alpha$ is present only for the charged $D^*$ and accounts for
Coulomb attraction in the final state \cite{Ginsberg:1968pz,
  Ginsberg:1969jh, Atwood:1989em}.
$\eta_{EW}$ arises from NLO box diagrams in which a photon or $Z$ is
exchanged together with the $W$ \cite{Sirlin:1981ie}.  
The kinematic factor $\chi(\omega)$
may be written
\begin{align}
\chi(\omega) = \frac{\omega + 1}{12}\left( 5\omega + 1 - \frac{8\omega(\omega - 1)r^*}{(1 - r^*)^2} \right).
\end{align}
The form factors $F_{D^{(*)}}(\omega)$ are related to hadronic matrix elements of the flavor-changing currents.  Compared to uncertainties in the form factors, uncertainties in the other quantities on the right-hand sides of Eqs.~\eqref{eq:b2d}, \eqref{eq:b2dstar} are small.

At zero-recoil, $F_{D^*}(1)=h_{A_1}(1)$, and only the axial current matrix
element contributes to the decay rate for \ds; heavy quark symmetry implies
$h_{A_1}(1)\approx 1$.

Given lattice QCD calculations of the form factors, experimental
measurements of the decay rates yield determinations of \vcb.  The
value of exclusive \vcb\ obtained in this way differs from the value
obtained from the inclusive decays $\bar{B}\to X_c\ell\bar{\nu}$ and
$\bar{B}\to X_s\gamma$ by 3.0$\sigma$.  This difference is correlated
with a 3.3$\sigma$ tension between \ek\ in the SM and
experiment~\cite{Bailey:2014qda}.  This tension vanishes when
inclusive \vcb\ is used to calculate \ek.

The form factors for the $\bar{B}\to D^{(*)}$ transition matrix elements are required
for calculations, in and beyond the SM, of the ratios $R(D^{(*)})\equiv
\mc{B}(\bar{B}\to D^{(*)}\tau\bar{\nu})/\mc{B}(\bar{B}\to
D^{(*)}\ell\bar{\nu})$, where $\ell=e,\mu$.  The BaBar Collaboration reported a
3.4$\sigma$ tension between its measurement of $(R(D),R(D^*))$ and the SM
value~\cite{Lees:2012xj}.  This tension is only partially relieved by a lattice
QCD calculation of $R(D)$~\cite{Bailey:2012jg}.  A lattice QCD calculation of
the form factors for $\bar{B}\to D^*$ at non-zero recoil, required for lattice
calculations of $R(D^*)$, does not (yet) exist.

The most precise determination of \vcb\ to date was obtained with the form
factor for \ds\ at zero recoil~\cite{Bailey:2014tva}.  Charm-quark
discretization effects dominate the uncertainty in the result.  One way to
reduce this systematic error is to generate data on finer lattices.  An
attractive alternative is to use a highly improved action.  Two actions are of
sufficient accuracy:  the highly improved staggered quark (HISQ) action and the
improved Fermilab action of Oktay and
Kronfeld~\cite{Follana:2006rc,Oktay:2008ex}.

The Oktay-Kronfeld (OK) action possesses the advantages of the
Sheikholeslami-Wohlert (SW) action with the Fermilab interpretation, including
control of discretization effects of fermions with arbitrary
mass~\cite{ElKhadra:1996mp}.  As $a\to 0$ or $m_Q\to\infty$, the discretization
effects vanish, while discretization effects of bottom quarks are smaller than
for charm quarks.  These features enable direct validation of bottom sector
calculations with calculations for the charm sector.

To reduce (heavy-quark) discretization effects in the axial and vector current
matrix elements, one must improve not only the action, but also the currents.
Following the work of Refs.~
\cite{ElKhadra:1996mp,Kronfeld:2000ck,Harada:2001fi,Harada:2001fj,Oktay:2008ex},
we introduce an improved field and are calculating quark-level matrix elements
to fix the coefficients of the higher order operators, as functions of the bare
quark masses.  In Sec.~\ref{sec:Psi} we briefly describe improvement for the
current and write down a complete set of operators that can appear in the
improved field, through third order in heavy quark effective theory (HQET).
Section \ref{sec:match} contains a description of our matching calculations.
In Sec.~\ref{sec:sum} we summarize completed and remaining work.

\section{\label{sec:Psi}Improved field}
%
%
The improvement program for fermions of arbitrary mass, in lattice units,
begins with the observation that time-space axis interchange symmetry, a
corollary of hypercubic rotation symmetry, is neither necessary nor convenient
for constructing actions closer to the renormalized trajectory.  Lifting
time-space axis interchange symmetry in the SW action, including only
higher-dimension operators that do not alter Wilson's time derivative, and
appropriately specifying the coefficients in the action as functions of the
fermion masses, one can systematically approach the renormalized trajectory
even though the fermion masses are large compared to the lattice
cutoff~\cite{ElKhadra:1996mp}.

Improvement of other operators, including the flavor-changing currents,
proceeds in much the same way.  In Ref.~\cite{ElKhadra:1996mp}, an improved
field, coinciding with the canonical Dirac field at tree-level and through
$\mc{O}(\bm{p})$, was introduced and shown to yield the desired continuum
matrix elements (at tree-level and through $\mc{O}(\bm{p})$).  Working at
tree-level and to higher order in the momentum expansion, an improved field
again suffices.  

The OK action is improved through $\mc{O}(\bm{p}^3)$ in HQET power
counting~\cite{Oktay:2008ex}.  Accordingly, we begin with an ansatz for the
improved field through $\mc{O}(\bm{p}^3)$,
\begin{align}
\Psi_I(x) &= e^{M_1/2} \biggl[ 1 + d_1 \bm{\gamma} \cdot \bm{D} + \tfrac{1}{2} d_2 \triangle^{(3)} 
+ \tfrac{1}{2} i d_B \bm{\Sigma} \cdot \bm{B} + \tfrac{1}{2} d_E \bm{\alpha} \cdot \bm{E} \CL
&+\ \tfrac{1}{4} d_{rE} \{ \bm{\gamma} \cdot \bm{D}, \bm{\alpha} \cdot \bm{E} \} + \tfrac{1}{4} d_{zE} \gamma_4 ( \bm{D} \cdot \bm{E} - \bm{E} \cdot \bm{D} ) 
+ \tfrac{1}{6} d_3 \gamma_i D_i \triangle_i + \tfrac{1}{2} d_4 \{ \bm{\gamma} \cdot \bm{D}, \triangle^{(3)} \} \CL
&+\ \tfrac{1}{4} d_5 \{ \bm{\gamma} \cdot \bm{D}, i \bm{\Sigma} \cdot \bm{B} \} + \tfrac{1}{4} d_{EE} \{ \gamma_4 D_4 , \bm{\alpha} \cdot \bm{E} \} 
+ \tfrac{1}{4} d_{z3} \bm{\gamma} \cdot ( \bm{D} \times \bm{B} + \bm{B} \times \bm{D} ) \biggr]\psi(x)\,,\label{eq:imp-field}
\end{align}
where the rest mass $M_1$ is related to the bare quark mass, $\bm{D}$ is the symmetric lattice covariant derivative, $\triangle_i$ is a covariant lattice second derivative, $\triangle^{(3)}$ is the three-dimensional lattice Laplacian, $\bm{\Sigma}$ and $\bm{\alpha}$ are $4\times 4$ matrices in spinor space, $\bm{B}$ and $\bm{E}$ are defined in terms of the clover field strength tensor, $\gamma_{\mu}$ are Euclidean gamma matrices, $\psi(x)$ is the unimproved field, and the coefficients $d_i$ depend on the mass of the fermion $\psi$ and other couplings in the action~\cite{ElKhadra:1996mp,Oktay:2008ex}.

The operators in this ansatz correspond to those in the OK action (bilinears)
through mass dimension 6, including operators whose coefficients vanish at
tree-level or are redundant.  Hence, all operators allowed by the lattice symmetries
are included.

The authors of Refs.~\cite{Harada:2001fi,Harada:2001fj} showed explicitly that
the currents constructed with the improved field through $\mc{O}(\bm{p})$,
including only the $d_1$ term introduced in Ref.~\cite{ElKhadra:1996mp},
coincides with that required to match through $\mc{O}(\Lambda_\mr{QCD}/m_Q)$ in
HQET.  To execute HQET matching, one enumerates operators in the lattice
current that correspond to those of the effective continuum HQET.  We have
begun the task of extending this enumeration to third order in HQET.

\section{\label{sec:match}Matching}
\begin{figure}[t!]
  \centering
  \vspace*{-5mm}
  \includegraphics{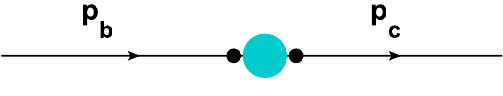}
  \caption{Diagram for quark-quark current matrix element for tree-level matching of the improved field.  The matching yields the parameters $d_i$ for $i=1,2,3,4$.}
  \label{fig:qJq}
\end{figure}

Here we consider matching conditions to determine the coefficients in
Eq.~\eqref{eq:imp-field}.  The authors of Ref.~\cite{ElKhadra:1996mp}
considered the quark-quark current matrix elements,
Fig.~\ref{fig:qJq}, with the flavor-changing current inserted between
external quarks.  They noted that at tree-level the difference between
lattice and continuum matrix elements is due to the difference between
the lattice and continuum spinors and spinor normalization factors.
Expanding the normalized continuum and lattice spinors through
$\mc{O}(\bm{p})$ and comparing the lattice and continuum matrix
elements yield $d_1$ (after equating the physical quark mass with the
kinetic quark mass).
Expanding the normalized continuum and lattice spinors through
$\mc{O}(\bm{p}^3)$, we find
\begin{align}
\sqrt{\frac{m_q}{E}}u(\xi,\bm{p}) &= \left[ 1 - \frac{i\bm{\gamma}\cdot\bm{p}}{2m_q} - \frac{\bm{p}^2}{8m_q^2} + \frac{ 3i ( \bm{\gamma} \cdot \bm{p} ) \bm{p}^2 }{16m_q^3} \right]u(\xi,\bm{0}) + \mc{O}(\bm{p}^4)\,, \\
\mc{N}(\bm{p})u^\mr{lat}(\xi,\bm{p}) &= e^{-M_1/2}\left[ 1 - \frac{i \zeta \bm{\gamma}\cdot\bm{p}}{2\sinh\,M_1} - \frac{\bm{p}^2}{8M_X^2} + \tfrac{1}{6} i w_3 \sum_{k=1}^3 \gamma_k p_k^3 + \frac{ 3i ( \bm{\gamma} \cdot \bm{p} ) \bm{p}^2 }{16M_Y^3} \right]u(\xi,\bm{0}) + \mc{O}(\bm{p}^4)\,,
\end{align}
where $\xi$ labels the linearly independent solutions, $E=\sqrt{m_q^2 + \bm{p}^2}$, 
the index $k$ is summed over the spatial directions,
\begin{align}
\frac{1}{M_X^2} &= \frac{\zeta^2}{\sinh^2\,M_1} + \frac{2r_s\zeta}{e^{M_1}}\,,\quad\quad
w_3 = \frac{3c_1 + \zeta/2}{\sinh\,M_1}\,,\\
\frac{1}{M_Y^3} &= \frac{8}{3\sinh\,M_1}\Biggl\{2c_2 + \frac{1}{4}e^{-M_1}\Biggl[\zeta^2 r_s (2 \coth\,M_1 + 1) \CL
&+\ \frac{\zeta^3}{\sinh\,M_1}\Biggl(\frac{e^{-M_1}}{2\sinh\,M_1} - 1 \Biggr) \Biggr] + \frac{\zeta^3}{4\sinh^2\,M_1} \Biggr\}\,,
\end{align}
$r_s$ is the coefficient of the spacelike Wilson term in the action, $\zeta
= \kappa_s/\kappa_t$, $c_{1,2}$ are coefficients in the mass-dimension 6
terms of the OK action, which modify the lattice spinors and mass shell {\it
via}
\begin{align}
K_i = \zeta \sin\,p_i\ \ \longrightarrow\ \ K_i = \sin\,p_i \left[\zeta - 2c_2\sum_{j=1}^3 (2\sin\,p_j/2)^2 - c_1(2\sin\,p_i/2)^2 \right]\,,
\end{align}
and the lattice spinor normalization factor for the OK action is
\begin{align}
\mc{N}(\bm{p}) &= \sqrt{\frac{\mu - \cosh\,E}{\mu \sinh\,E}}
\quad\text{where}\quad 
\cosh\,E = \frac{1 + \mu^2 + \bm{K}^2}{2\mu}\,, 
\quad\text{and}\quad \bm{K}^2 = \sum_i K_i^2\\
\mu &= 1 + m_0 + \tfrac{1}{2}r_s\zeta\sum_{i=1}^3 (2\sin\,p_i/2)^2\,,
\end{align}
where $m_0$ is the bare heavy quark mass.
As the lattice spacing tends to zero, the masses $M_{X,Y}$ tend to the rest
mass $M_1$.  With the OK action (matched at tree-level), the rotation breaking
parameter $w_3 = c_B = r_s$.

The mismatch between the lattice and continuum normalized spinors is
compensated by the rotation parameters $d_i$ in the current constructed from
the improved field.

At tree-level the terms in Eq.~\eqref{eq:imp-field} with chromoelectric and
chromomagnetic fields do not contribute to the matrix elements of
Fig.~\ref{fig:qJq}; only the terms with coefficients $d_{1,2,3,4}$ contribute.
Setting the gauge links to one in the covariant derivatives, we note the
additional factors entering contractions between differentiated fields and
external quark states,
\begin{align}
\psi\to \triangle^{(3)}\psi &\ \ \text{leads to}\ \ u^\mr{lat}\to -\sum_i(2\sin\,p_i/2)^2 u^\mr{lat}\,,\\
\psi\to D_i\triangle_i\psi &\ \ \text{leads to}\ \ u^\mr{lat}\to -i\sin\,p_i(2\sin\,p_i/2)^2 u^\mr{lat}\,,\\
\psi\to D_i\triangle^{(3)}\psi &\ \ \text{leads to}\ \ u^\mr{lat}\to -i\sin\,p_i\sum_j(2\sin\,p_j/2)^2 u^\mr{lat}\,.
\end{align}

Then calculating the continuum and lattice matrix elements and demanding equality through $\mc{O}(\bm{p}^3)$,
\begin{align}
\sqrt{\frac{m_c}{E_c}}  \bar{u}_c(\xi_c,\bm{p}_c) 
\Gamma  \sqrt{\frac{m_b}{E_b}}  u_b(\xi_b,\bm{p}_b)
=
\mc{N}_c(\bm{p}_c)  \bar{u}^\mr{lat}_c(\xi_c,\bm{p}_c) \bar{R}(\bm{p}_c)
\Gamma  \mc{N}_b(\bm{p}_b) R(\bm{p}_b) u^\mr{lat}_b(\xi_b,\bm{p}_b)\,,\label{eq:demand}
\end{align}
where $R(\bm{p})$ represents contributions from the improvement terms 
in Eq.~\eqref{eq:imp-field} and depends on the parameters $d_i$.

We find matching the $\mc{O}(\bm{p})$ terms yields $d_1$, matching the $\mc{O}(\bm{p}^2)$ terms yields $d_2$, matching the rotation breaking terms yields $d_3$, and matching the rotation invariant terms of $\mc{O}(\bm{p}^3)$ yields $d_4$.  We have
\begin{align}
d_1 &= \frac{\zeta}{2\sinh\,M_1} - \frac{1}{2m_q}\,,\label{eq:d1}\\
d_2 &= d_1^2 - \frac{r_s\zeta}{2e^{M_1}}\,,\label{eq:d2}\\
d_3 &= -d_1 + w_3\,,\label{eq:d3}\\
d_4 &= -\frac{d_1}{8M_X^2} + \frac{d_2\zeta}{4\sinh\,M_1}+ \frac{3}{16}\left(\frac{1}{M_Y^3} - \frac{1}{m_q^3} \right)\,,\label{eq:d4}
\end{align}
where $m_q$ is to be taken equal to the kinetic mass $M_2$.

The results for $d_{1,2}$ are contained already in Ref.~\cite{ElKhadra:1996mp}.

To specify the remaining coefficients in Eq.~\eqref{eq:imp-field}, we are considering four-quark current matrix elements.  For example,
\begin{align}
\matrixe{q^f(\xi_2,\bm{p}_2)\,c(\xi_c,\bm{p}_c)}{J^{cb}_{\Gamma}}{b(\xi_b,\bm{p}_b)q^g(\xi_1,\bm{p}_1)}\,,
\end{align}
where $f,g$ are flavor indices and $J^{cb}_{\Gamma}$ is the bottom ($b$) to charm ($c$) flavor-changing current.

\begin{figure}[t!]
  \centering
  \subfloat[\label{sfig:a}]{%
    \includegraphics[width=0.4\textwidth]{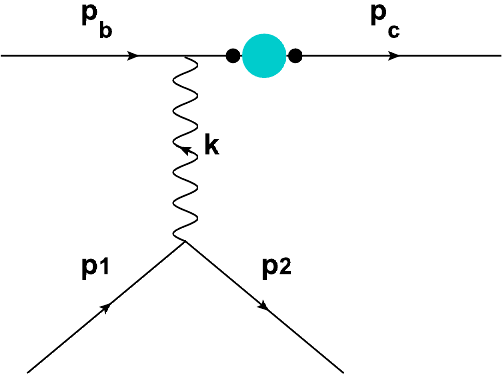}
  }\hspace{0.5cm}
  \subfloat[\label{sfig:b}]{%
    \includegraphics[width=0.4\textwidth]{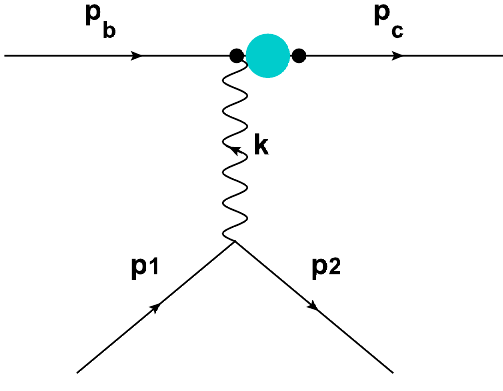}
  }\\
  \subfloat[\label{sfig:c}]{%
    \includegraphics[width=0.4\textwidth]{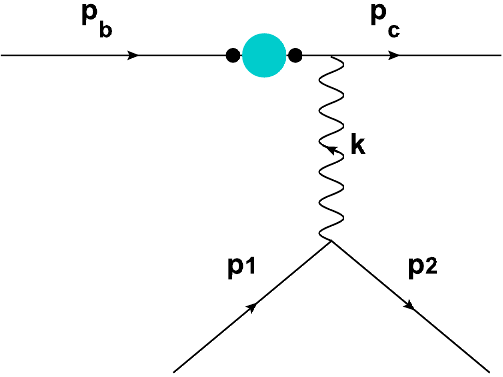}
  }\hspace{0.5cm}
  \subfloat[\label{sfig:d}]{%
    \includegraphics[width=0.4\textwidth]{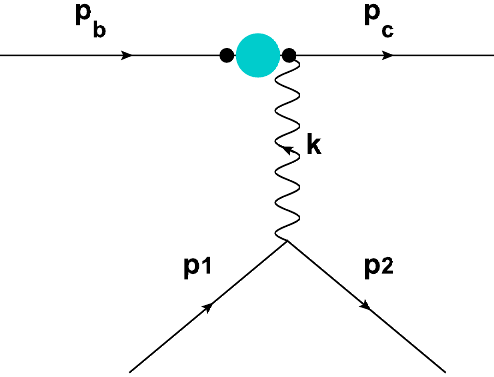}
  }
  \caption{Diagrams for four-quark current matrix elements for tree-level matching of the improved field.  The large circles are electroweak current insertions, and the small circles represent the field improvement terms.
}
\label{fig:qqJqq}
\end{figure}

Tree-level diagrams contributing to this matrix element are shown in
Fig.~\ref{fig:qqJqq}.  The large filled circle is the current, and the small
circles are the improvement terms.  Gluon exchange can occur with the external
bottom or charm quark or with the higher order operators in the improved bottom
or charm field.  

The OK action coefficients are specified (in part) through matching the
quark-gluon vertex~\cite{Oktay:2008ex}.  Turning to the current-gluon vertices
and exchange with the external bottom and charm quarks, we expand in the
external momenta.  From the spatial component of the gluon vertex at lowest
order, we recover the result for $d_1$, and we expect results for $d_{2,B}$ in
agreement with Ref.~\cite{ElKhadra:1996mp}.  From the time component and higher
order terms in the momentum expansion, we anticipate information about the
parameters $d_E,\ d_{rE},\ d_{zE},\ d_5,\ d_{EE},$ and $d_{z3}$.  

The rotation parameters $d_1, d_2, d_B$, and $d_E$ can also be obtained from
the Foldy-Wouthuysen-Tani transformed field by including operators of
$\mathcal{O}(\bm{p}^2)$ and matching the Hamiltonian \cite{ElKhadra:1996mp}.
We note that $d_E$ cannot be altered by the higher order terms of
$\mathcal{O}(\bm{p}^3)$ appearing in the OK action.

\section{\label{sec:sum}Summary}
To reduce the discretization effects of heavy quarks, the OK action is improved
through third order in HQET power counting~\cite{Oktay:2008ex}.  Systematic
improvement of the hadronic $\bar{B}\to D^{(*)}$ matrix elements needed to extract
\vcb\ from the branching fractions of \dps\ decays requires improvement of the
(axial and vector) $b\to c$ currents through the same order in HQET.

The authors of Ref.~\cite{ElKhadra:1996mp,Harada:2001fi,Harada:2001fj} showed
that currents constructed from an improved quark field suffice for improvement
of the hadronic matrix elements through $\mc{O}(\bm{p})$, or first order in
HQET.  The field improvement operators shown in Eq.~\eqref{eq:imp-field} suffice
for improvement of quark-quark current matrix elements, cf.
Eq.~\eqref{eq:demand}.  Matching to the continuum matrix elements yields the
parameters $d_{1,2,3,4}$, Eqs.~\eqref{eq:d1}, \eqref{eq:d2}, \eqref{eq:d3}, and
\eqref{eq:d4}.  

To specify the coefficients of the remaining operators, we are matching
four-quark current matrix elements.  To demonstrate improvement through third
order in HQET, we are enumerating operators in the lattice currents and the
(effective continuum) currents of HQET~\cite{Harada:2001fi,Harada:2001fj}.

\acknowledgments
J.A.B. is supported by the Basic Science Research Program of the National
Research Foundation of Korea (NRF) funded by the Ministry of Education
(No.~2014027937).  The research of W.~Lee is supported by the Creative Research
Initiatives Program (No.~2014001852) of the NRF grant funded by the Korean
government (MEST).  
W. Lee would like to acknowledge support from KISTI supercomputing center
through the strategic support program for the supercomputing application
research [No.~KSC-2013-G2-005].
\bibliography{refs}

\providecommand{\href}[2]{#2}\begingroup\raggedright\begin{thebibliography}{10}

\bibitem{Manohar:2000dt}
A.~V. Manohar and M.~B. Wise, {\it {Heavy quark physics}},  {\em Cambridge}
  (2000).

\bibitem{Sirlin:1981ie}
A.~Sirlin {\em Nucl.Phys.} {\bf B196} (1982) 83.

\bibitem{Ginsberg:1968pz}
E.~S. Ginsberg {\em Phys.Rev.} {\bf 171} (1968) 1675.

\bibitem{Ginsberg:1969jh}
E.~S. Ginsberg {\em Phys.Rev.} {\bf 162} (1967) 1570.

\bibitem{Atwood:1989em}
D.~Atwood and W.~J. Marciano {\em Phys.Rev.} {\bf D41} (1990) 1736.

\bibitem{Bailey:2014qda}
J.~A. Bailey, Y.-C. Jang, and W.~Lee {\em PoS} {\bf LATTICE2014} (2014) 371,
  [\href{http://xxx.lanl.gov/abs/1410.6995}{{\tt 1410.6995}}].

\bibitem{Lees:2012xj}
J.~Lees {\em et~al.} {\em Phys.Rev.Lett.} {\bf 109} (2012) 101802,
  [\href{http://xxx.lanl.gov/abs/1205.5442}{{\tt 1205.5442}}].

\bibitem{Bailey:2012jg}
J.~A. Bailey and others [FNAL/MILC] {\em Phys.Rev.Lett.} {\bf 109} (2012)
  071802, [\href{http://xxx.lanl.gov/abs/1206.4992}{{\tt 1206.4992}}].

\bibitem{Bailey:2014tva}
J.~A. Bailey and others [FNAL/MILC] {\em Phys.Rev.} {\bf D89} (2014) 114504,
  [\href{http://xxx.lanl.gov/abs/1403.0635}{{\tt 1403.0635}}].

\bibitem{Follana:2006rc}
E.~Follana and others [HPQCD/UKQCD] {\em Phys.Rev.} {\bf D75} (2007) 054502,
  [\href{http://xxx.lanl.gov/abs/hep-lat/0610092}{{\tt hep-lat/0610092}}].

\bibitem{Oktay:2008ex}
M.~B. Oktay and A.~S. Kronfeld {\em Phys.Rev.} {\bf D78} (2008) 014504,
  [\href{http://xxx.lanl.gov/abs/0803.0523}{{\tt 0803.0523}}].

\bibitem{ElKhadra:1996mp}
A.~X. El-Khadra, A.~S. Kronfeld, and P.~B. Mackenzie {\em Phys.Rev.} {\bf D55}
  (1997) 3933--3957, [\href{http://xxx.lanl.gov/abs/hep-lat/9604004}{{\tt
  hep-lat/9604004}}].

\bibitem{Kronfeld:2000ck}
A.~S. Kronfeld {\em Phys.Rev.} {\bf D62} (2000) 014505,
  [\href{http://xxx.lanl.gov/abs/hep-lat/0002008}{{\tt hep-lat/0002008}}].

\bibitem{Harada:2001fi}
J.~Harada, S.~Hashimoto, K.-I. Ishikawa, A.~S. Kronfeld, T.~Onogi, {\em et~al.}
  {\em Phys.Rev.} {\bf D65} (2002) 094513,
  [\href{http://xxx.lanl.gov/abs/hep-lat/0112044}{{\tt hep-lat/0112044}}].

\bibitem{Harada:2001fj}
J.~Harada, S.~Hashimoto, A.~S. Kronfeld, and T.~Onogi {\em Phys.Rev.} {\bf D65}
  (2002) 094514, [\href{http://xxx.lanl.gov/abs/hep-lat/0112045}{{\tt
  hep-lat/0112045}}].

\end{thebibliography}\endgroup
\end{document}